\documentclass[%
 reprint,
 superscriptaddress,
 amsmath,amssymb,
 aps,
]{revtex4-2}

\usepackage{graphicx}
\usepackage{dcolumn}
\usepackage{bm}
\usepackage{hyperref}

\begin{document}

\title{Crystal Toolkit: A Web App Framework to Improve Usability and Accessibility of Materials Science Research Algorithms}



\author{Matthew Horton}
\email{mkhorton@lbl.gov}
\affiliation{Materials Science Division, Lawrence Berkeley National Lab}
\affiliation{Department of Materials Science and Engineering, University of California, Berkeley}



\author{Jimmy-Xuan Shen}
\affiliation{Materials Science Division, Lawrence Berkeley National Lab}
\affiliation{Department of Materials Science and Engineering, University of California, Berkeley}

\author{Jordan Burns}
\affiliation{Department of Materials Science and Engineering, University of California, Berkeley}
\affiliation{Energy Storage and Distributed Resources Division, Lawrence Berkeley National Lab}

\author{Orion Cohen}
\affiliation{Materials Science Division, Lawrence Berkeley National Lab}
\affiliation{Department of Chemistry, University of California, Berkeley}

\author{Fran\c{c}ois Chabbey}
\affiliation{Materials Science Division, Lawrence Berkeley National Lab}

\author{Alex M. Ganose}
\affiliation{Department of Chemistry, Imperial College London} 

\author{Rishabh Guha}  
\affiliation{Materials Science Division, Lawrence Berkeley National Lab} 

\author{Patrick Huck}
\affiliation{Materials Science Division, Lawrence Berkeley National Lab} 

\author{Haoming Howard Li}  
\affiliation{Department of Materials Science and Engineering, University of California, Berkeley}
\affiliation{Materials Science Division, Lawrence Berkeley National Lab} 

\author{Matthew McDermott}
\affiliation{Department of Materials Science and Engineering, University of California, Berkeley}
\affiliation{Materials Science Division, Lawrence Berkeley National Lab} 

\author{Joseph Montoya}
\affiliation{Toyota Research Institute}

\author{Guy Moore}
\affiliation{Department of Materials Science and Engineering, University of California, Berkeley}
\affiliation{Materials Science Division, Lawrence Berkeley National Lab} 

\author{Jason Munro}
\affiliation{Materials Science Division, Lawrence Berkeley National Lab} 

\author{Cody O'Donnell}
\affiliation{Materials Science Division, Lawrence Berkeley National Lab}

\author{Colin Ophus}  
\affiliation{The National Center for Electron Microscopy, Molecular Foundry, Lawrence Berkeley National Lab}  

\author{Guido Petretto}
\affiliation{Matgenix SRL}
\affiliation{Modelling Division, Institute of Condensed Matter and Nanosciences, Université catholique de Louvain} 

\author{Janosh Riebesell}  
\affiliation{Cavendish Laboratory, University of Cambridge}
\affiliation{Molecular Foundry, Lawrence Berkeley National Lab}  

\author{Steven Weitzner}
\affiliation{Lawrence Livermore National Laboratory}

\author{Brook Wander}  
\affiliation{Department of Chemical Engineering, Carnegie Mellon University}

\author{Donald Winston}
\affiliation{Energy Technologies Area, Lawrence Berkeley National Lab} 

\author{Ruoxi Yang}
\affiliation{Materials Science Division, Lawrence Berkeley National Lab}

\author{Steven Zeltmann} 
\affiliation{Department of Materials Science and Engineering, University of California, Berkeley} 

\author{Anubhav Jain}
\affiliation{Materials Science Division, Lawrence Berkeley National Lab} 

\author{Kristin A. Persson}
\email{kapersson@lbl.gov}
\affiliation{Molecular Foundry, Lawrence Berkeley National Lab}
\affiliation{Department of Materials Science and Engineering, University of California, Berkeley}


\begin{abstract}
Crystal Toolkit is an open source tool for viewing, analyzing and transforming crystal structures, molecules and other common forms of materials science data in an interactive way. It is intended to help beginners rapidly develop web-based apps to explore their own data or to help developers make their research algorithms accessible to a broader audience of scientists who might not have any training in computer programming and who would benefit from graphical interfaces. Crystal Toolkit comes with a library of ready-made components that can be assembled to make complex web apps: simulation of powder and single crystalline diffraction patterns, convex hull phase diagrams, Pourbaix diagrams, electronic band structures, analysis of local chemical environments and symmetry, and more. Crystal Toolkit is now powering the Materials Project website frontend, providing user-friendly access to its database of computed materials properties. In the future, it is hoped that new visualizations might be prototyped using Crystal Toolkit to help explore new forms of data being generated by the materials science community, and that this in turn can help new materials scientists develop intuition for how their data behaves and the insights that might be found within.  Crystal Toolkit will remain a work-in-progress and is open to contributions from the community.
\end{abstract}

\maketitle

\section{Introduction\protect\\}

Effective communication of scientific results is becoming ever more essential: an exponential increase in quantity of papers produced has lead to an ``attention decay''\cite{attentiondecay}, whereby it is becoming increasingly difficult for researchers to stay abreast of the published literature. Furthermore, new scientific techniques are generating a vast increase in the quantity of data: this, too, can present a problem, as insights are hidden in these large data sets, which are often difficult to explore. This is a challenge to our community and also an opportunity to find new solutions for better scientific communication. This paper describes improvement in this area by the introduction of a new tool for computational materials scientists to help them explore and share their data, and increase the impact of their work by making it easier to understand and build connections between data sets.

\begin{figure*}
    \includegraphics[width=6in]{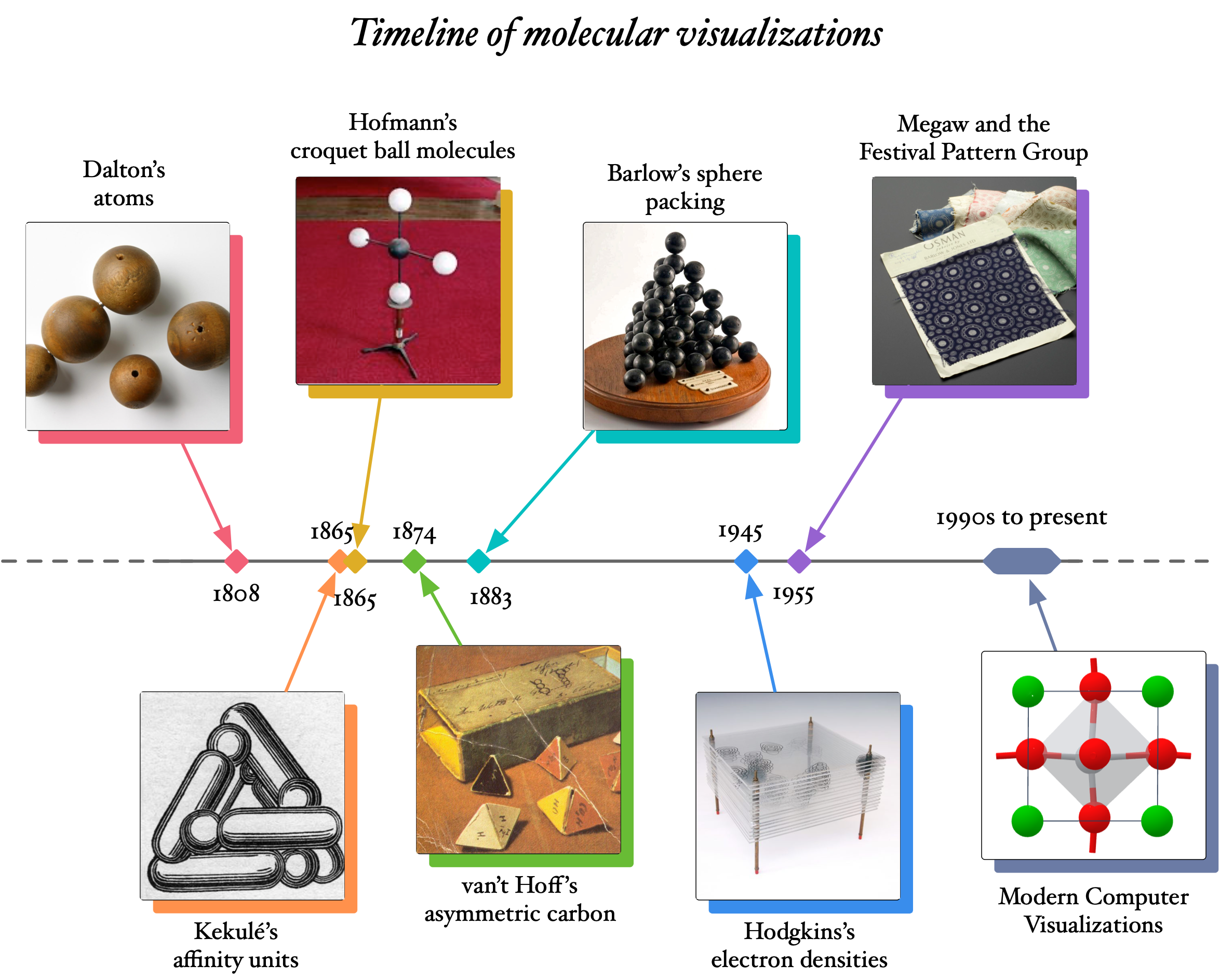}
    \caption{A timeline of molecular and crystallographic visualizations with several key advances highlighted. Throughout the history of chemistry and materials science, effective visualization has proven an invaluable tool to help develop intuition and unlock key new insights that have advanced the field.}
    \label{fig:timeline}
\end{figure*}

Materials science and chemistry have always benefited from a visual exploration of concepts, whereby visual representations have not only been used as a tool to accelerate learning but have actually enabled new scientific discoveries\cite{croquetballs} (see Figure \ref{fig:timeline}). These include Dalton's atoms at the advent of atomic theory, Kekul\'e's affinity units (initially a convenient orthographic device but that led to the discovery of the benzene ring), Hofmann's molecules built from croquet balls (introducing a strong sense of the spatial arrangement of molecules, and colour schemes for elements that are still in use today), van't Hoff's asymmetric carbon and the concept of stereoisomers, Barlow and the understanding of close packing in crystal structures, Hodgkin and the visualization of electron charge densities in real-space, and even Megaw's popularization of materials science through the Festival Pattern Group--the list goes on! These examples demonstrate how effective visualization has led to essential leaps throughout the history of materials science, and justifies close attention to this area -- for education\cite{kobayashi2021virtual,lehtola2022free} as well as research.


In the modern era, there has been an explosion of options for molecular visualizations enabled by advances in computation which support new computational simulation techniques. A full review of these would be a substantial effort, but include XCrysDen\cite{kokalj_xcrysdennew_1999}, OVITO\cite{stukowski_visualization_2010}, VESTA\cite{momma_vesta_2011}, Avogadro\cite{hanwell_avogadro:_2012}, VMD\cite{humphrey_vmd:_1996}, Jmol\cite{hanson_jmol_2010}, 3DMol\cite{rego_3dmol.js:_2015}, nglviewer\cite{rose_ngl_2015}, AtomEye\cite{li_atomeye:_2003}. Each of these tools has developed and refined what is possible in molecular visualization, and have, in turn, each contributed to scientific insights. It is reasonable therefore to consider that the problem of molecular visualization might be solved and ask what benefit yet another option might bring. 

Crystal Toolkit has been developed to address new, specific requirements emerging from the computational materials science community. It provides not only new means of molecular and crystal visualization, but goes beyond these and facilitates the visualization of additional forms of data relevant to the field. These requirements come specifically as a result of the rise of high-throughput computational approaches and the vast quantities of data that they create, as well as from recent efforts which closely link computation with experiment to aid in the design of new materials.


\section{Design Goals}

\begin{figure}
    \centering
    \includegraphics[width=0.5\textwidth]{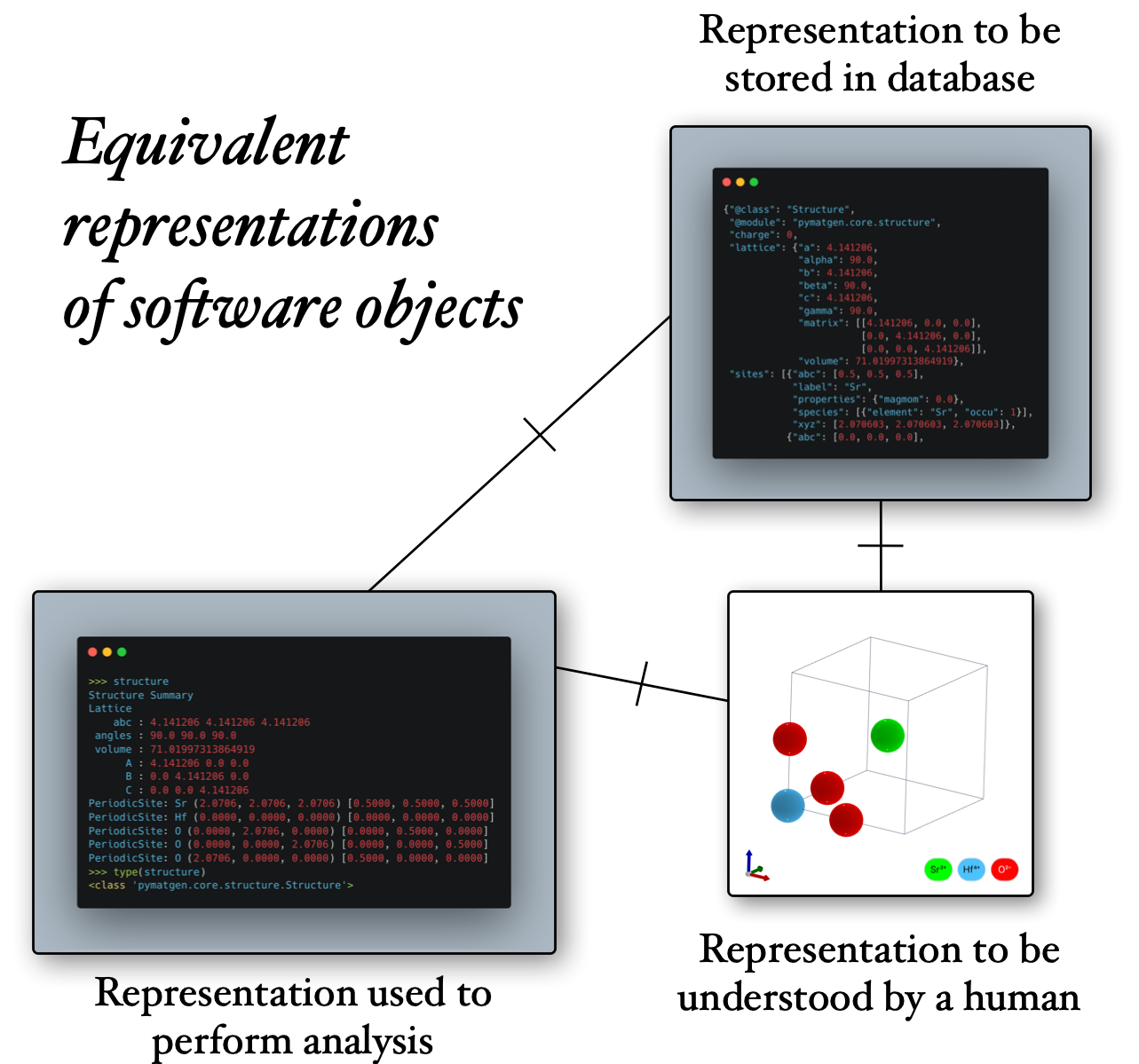}
    \caption{A design goal of Crystal Toolkit is to provide human-centered representations of software objects, and specifically to ensure that the default representations are as faithful to the underlying software object as possible so as to help the user gain intuition about the use of these objects.}
    \label{fig:representations}
\end{figure}

\begin{figure*}
    \centering
    \includegraphics[width=6in]{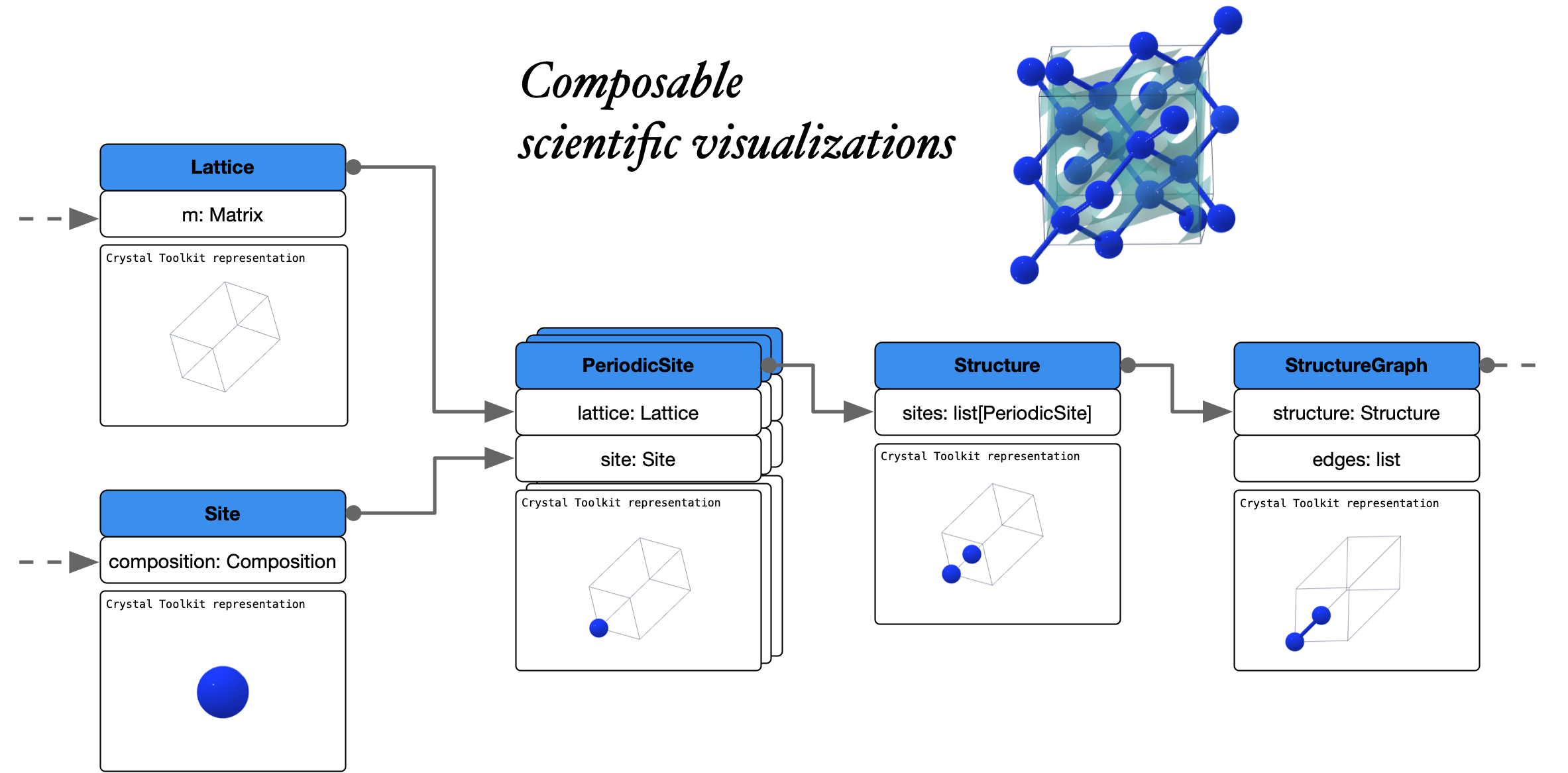} 
    \caption{In a similar way that object-orientated programming allows complex objects to be constructed and composed from simpler objects, visualizations in Crystal Toolkit are designed to map directly to the underlying software object and to be composable, so that complex visualizations can easily be built up. This example shows part of how a visualization for silicon is constructed, with the inset figure showing the final visualization after transformation to a conventional cell and addition of an electron charge density layer.}
    \label{fig:composable}
\end{figure*}

\begin{figure*}
    \centering
    \includegraphics[width=6in]{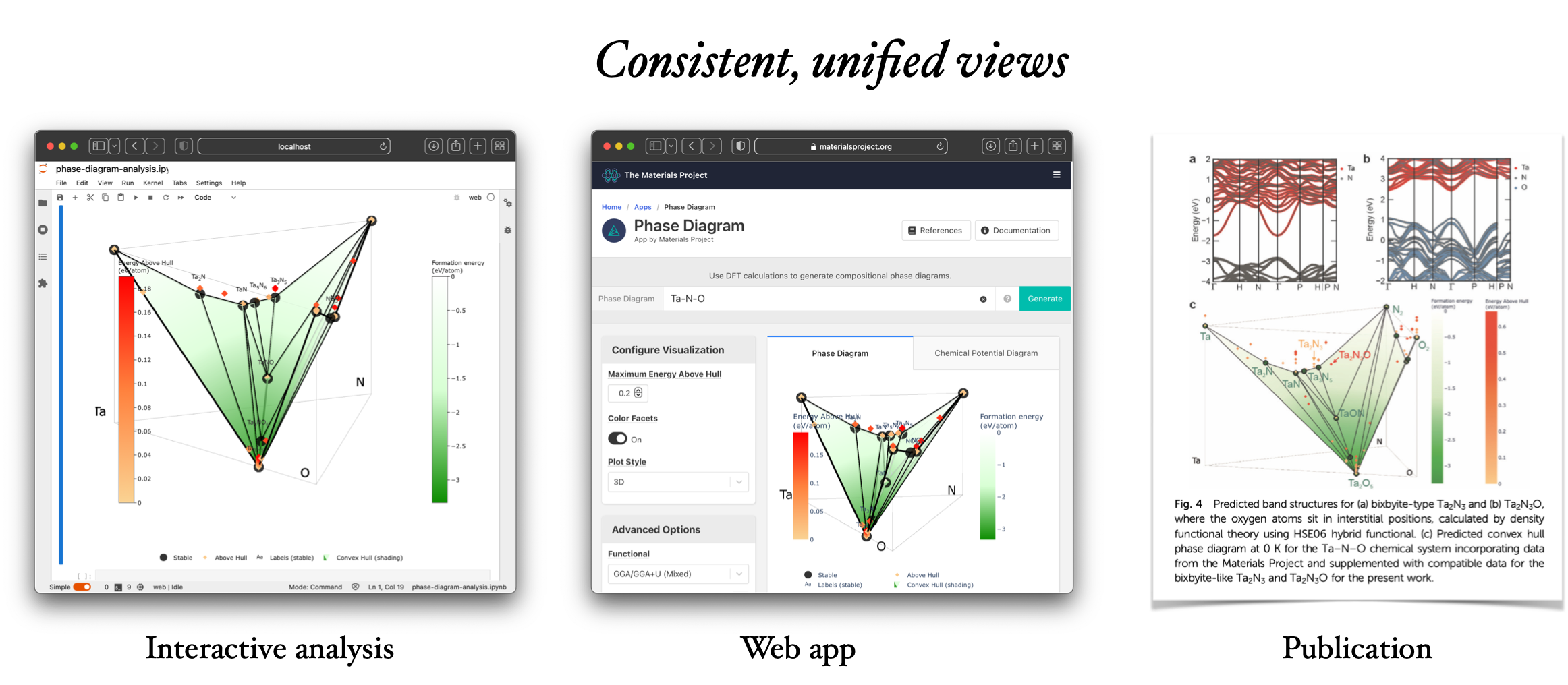} 
    \caption{A significant motivation for the design of Crystal Toolkit is to allow consistent views of the same information across journal publications, web apps, and also during initial data generation and analysis, so as to reduce the opportunity for loss of information or introduction of errors as the data moves from one location to another. Publication figure is taken from Jiang \textit{et al.}\cite{jiang2021metastable}}
    \label{fig:views}
\end{figure*}

Several key goals have governed the design of Crystal Toolkit, informed by historical and technological context, as well as the current needs of the computational materials science community. These include:

\begin{itemize}
    \item \textbf{A human-centered approach.} Since the early study of ``human-computer interaction''\cite{hci}, people have framed the relationship between human and computer as a dialogue. Unfortunately, this dialogue is often too focused on the computer, and humans are forced to adapt to the machine rather than the inverse. In scientific research, this often manifests as poorly-documented codes which are difficult to use or compile, and has had a tremendous cost on scientific advance: in the best case, it delays progress of new users by months or years as they have to learn the quirks of these computational tools, while in the worst case it can alienate new scientists and drive them out of the field entirely, or even result in the publication of incorrect results due to misunderstanding and misinterpretation of a computer code or its output. It is critical that the next generation of tools that we develop as a community start with a human-centered approach to avoid these outcomes.
    \item \textbf{Strong mapping of visualizations to specific software objects.} Individual software objects (for example, a \texttt{Structure} object representing a crystal structure) are often difficult to introspect and understand what information they contain. If you are a proficient programmer you can inspect its attributes or read the relevant source code, but this is often inaccessible to new learners. Visualizations allow both novice and experienced users to leverage the brain's immense capacity for visual information processing\cite{franconeri2021}. Visual representations of data structures might include a formal representation such as a UML (Unified Modeling Language) diagram, but might more usefully be a 3D visualization or a graphical plot. Crystal Toolkit aims to provide understandable visualizations for the commonly used software objects used by the materials science community.
    \item \textbf{Faithful representations of software objects.} It is important that a visualization faithfully represents an underlying software object, and that this in turn is a faithful representation of the corresponding scientific abstraction (see Figure \ref{fig:representations}). These two degrees of separation--between concept and code, and code and visualization--are fraught, and can introduce problems. For a specific example, some molecular visualizations will automatically detect and insert bonds that are not present in the underlying software object. While useful, it impedes understanding of the object, and can introduce confusion for new developers. A better option is to only visualize the information actually present in the object \textit{by default} with the option of including additional ornaments.
    \item \textbf{Composability of visualizations.} Object-orientated programming is a common pattern in modern programming, whereby complex objects can inherit from and be composed from simpler objects (see Figure \ref{fig:composable}). Where possible, if a complex object is built from a set of simpler objects, and these objects in turn already have visualizations available, then it should be possible that the visualization of the more complex object can be composed from these already existing visualizations. This allows for reuse and experimentation, and helps enforce the faithfulness of the representation.
    \item \textbf{Empowering users of all experience levels.} Many scientists do not receive formal training in programming, or might receive training in an informal and ad-hoc manner. Nevertheless, ability in programming is becoming increasingly essential to perform scientific analysis. Crystal Toolkit aims to tackle this issue in two ways, by (1) allowing complex web apps to be constructed such that the latest computational materials science algorithms can be used directly via a web app, without any additional programming required by the user, and (2) by providing a roadmap for scientists with some programming training to make their own web apps to explore their own data. To achieve (2), Crystal Toolkit should allow apps to be constructed and prototyped from just a few lines of Python code. Python was chosen since it is now the \textit{de facto} standard programming language used for a wide range of scientific tasks, so a language that a new scientist is most likely to have experience with.
    \item \textbf{Portability of representations between different media.} Information can be lost when converting from one representation of a given object to another. For example, a scientist might use one form of plotting or visualization while performing analysis, use another for a publication, and a third for interactive use on a website (see Figure \ref{fig:views}). Ideally, all of these views of the same object should be the same, within the limitations of the respective medium. This reduces the risk of loss of information and also reduces the effort required to share data in different places; for example, by making it easier to offer an interactive version of a visualization included in a publication. 
    \item \textbf{An open, inclusive attitude to new contributors.} The Crystal Toolkit authors want the framework to live and develop over time, and hope new developers will join in this process. A Zenodo citation will also be provided to ensure new developers can be acknowledged for their contributions, and all users of Crystal Toolkit are encouraged to cite both this design manuscript and also the Zenodo citation.
    \item \textbf{A tool for learning and developing intuition.} In science, we are familiar with how to think about the correspondence between reality and our models of reality, where they agree and are helpful, and where the limits of our models can obscure truth (for example, imagining a bond as a spring). However, with the rise of computational science, we are less practiced in examining our simulations, their predictions and our visual or textual representations of them, whether in the form of graphics, animations or graphs. A good visualization tool should help develop intuition for the models we use, and act as an educational aid.
\end{itemize}

\section{A Library of Ready-to-Use Components}

Crystal Toolkit comes with several components included and ready to use. These components map to existing data types used in the materials science community, and mainly to specific classes available in the \textit{pymatgen}\cite{ong2013python} code. These components can be used individually, or can be linked together; for example, a ``symmetry'' component can be linked to a ``crystal structure'' component so that symmetry information is always shown for the same crystal structure being visualized.

The following is only an initial list of components, with components expected to evolve as development proceeds and new components are made available, so the current list should be considered a snapshot of Crystal Toolkit's capabilities at the time of publication. For an up-to-date list, please consult the Crystal Toolkit documentation\cite{docs}. See Figure \ref{fig:library}, left panel, for illustrations of several of these components.

\begin{enumerate}
    \item \textbf{Crystal Structures.} Visualize crystal structures with a variety of visualization options, including different colour schemes (including accessible colour schemes), choice of atom size and bonding algorithms\cite{pan2021benchmarking}. The structure visualizer also allows transformation to different crystallographic settings and downloading of crystal structures to common file formats including CIF. Maps to \texttt{Structure} and \texttt{StructureGraph} objects.
    \item \textbf{Molecules.} Visualize molecules in a similar way to crystal structures. Maps to \texttt{Molecule} and \texttt{MoleculeGraph} objects.
    \item \textbf{Band Structures.} Visualize electronic or phonon band structures and densities of states. Interact with the band structure by zooming in to specific regions or, in future, by selecting different k-path conventions\cite{munro2020improved}.  Maps to \texttt{BandStructure} and \texttt{DOS} objects. 
    \item \textbf{Pourbaix Diagrams.} Visualize Pourbaix (aqueous stability) diagrams built from experimental and theoretical data. Interact by changing ion concentrations, or show a heatmap of free energy for a specific material across voltage-pH space. Maps to \texttt{PourbaixDiagram} objects.
    \item \textbf{X-ray Absorption Spectra.} Visualize X-ray absorption spectra, or other kinds of spectra, and interact by applying Gaussian broadening. Maps to \texttt{Spectrum} objects.
    \item \textbf{Phase Diagrams.} Visualize phase diagrams generated with a convex hull construction. Supports 1-, 2-, 3- and 4-element phase diagrams, with 3-element phase diagrams viewable in both 2D and 3D with $z$-axis of formation enthalpy, and 4-element phase diagrams in 3D. Viewing in chemical potential space is also possible. Interaction includes transformation into a grand potential phase diagram by inclusion of open elements. Maps to \texttt{PhaseDiagram} objects.
    \item \textbf{Symmetry.} Show symmetry information, including space group, point group and detected Wyckoff labels, for a given crystal structure, based on spglib\cite{togo2018texttt}. Interaction includes being able to change tolerances for detection of symmetry, an important consideration for computationally-obtained crystal structures. Improvements include integration with Auguste\cite{larsen2020minimum} to show minimum strain values.
    \item \textbf{Diffraction Patterns.} This component accepts a crystal structure and can generate either an X-ray powder diffraction pattern or a transmission electron microscope single crystal diffraction pattern. In both cases, the input crystal structure is first transformed into its conventional setting such that the peaks and diffraction spots respectively are annotated with the conventional lattice indices. Interaction includes the ability to apply Scherrer broadening to estimate finite size effects.
    \item \textbf{Wulff Shapes.} Visualize a Wulff shape and interact to examine specific facets and surface energies. Maps to \texttt{WulffShape} objects. In an app, it is recommended to link this component to a crystal structure component to allow corresponding surfaces to be visualized on interaction with a given facet.
    \item \textbf{Molecular Graphs.} Molecular or crystal graphs, representing atoms as nodes and bonds (or other information) as edges, can be represented as interactive force-directed graphs. Maps to \texttt{StructureGraph} or \texttt{MoleculeGraph} objects.
    \item \textbf{Periodic Table.} This component provides a means to allow selection of specific elements, to be used as an input for another component, and also supports displaying a heatmap for a set of data associated with a list of elements.
    \item \textbf{Brillouin Zones.} Visualize Brillouin zones and k-paths in 3D. Improvements include linking to band structure components to allow a given high-symmetry line in a band structure diagram to be highlighted in the corresponding Brillouin zone. Plotting of Fermi surfaces in the Brillouin zone is also possible using IFermi\cite{ganose2021ifermi}.
    \item \textbf{Transformations Component.} This component accepts a crystal structure as input and provides a means to transform that crystal structure using \textit{pymatgen}. Maps to \texttt{Structure} and \texttt{AbstractTransformation} objects.
\end{enumerate}

\section{Open-Source Development Model}

Crystal Toolkit is developed as Open Source Software under a BSD license\cite{bsd}, with the intent to encourage involvement of as many interested scientists as possible in its development. Discussions on features and bugs are conducted openly, currently on GitHub, and are open for participation from anyone.

This approach is motivated by the belief that scientific tools should be open to allow their results to be reproducible, and to try to avoid issues of older tools being abandoned as developers move on and then, over time, ceasing to function. It is essential that new tools are developed with sustainability in mind, so that they can continue to work into the future as the development team or lead maintainers change.



The technical implementation of Crystal Toolkit is expected to change over time as new technology becomes available. Thus, the current manuscript is primarily focused on the design goals and overall philosophy of the project.

At time of writing, the project uses both Python code and JavaScript code, and is built upon the Dash framework\cite{plotly} by Plotly, with additional custom components written using React\cite{mpreact} and packaged as Dash components\cite{dashmp}. React is a JavaScript library for creating modern web apps that has now seen extensive use by the web development community. 

Dash was chosen specifically because it allowed an interactive web app to be written in just a single file in the Python programming language yet provide common web developer convenience features such as hot-module reloading on file changes. This enables apps to be prototyped and developed more rapidly and helps catch bugs more quickly compared to having to write code ``in the dark'' without immediate feedback. All web components, including custom React components written in JavaScript, map back to automatically-generated and documented Python classes, such that they can be used by Python developers without prior web development experience. However, the ability to write new, custom React components empowers dedicated web developers to extend the functionality of this code in useful ways (e.g., the development of the Periodic Table component). Such a parallel development model enables web experts to extend the code while materials science domain experts can apply and use the code, without either blocking the other's work. Furthermore, since no client state is stored on the server by default, Dash readily scales horizontally and is appropriate even for web apps which experience a large amount of traffic like the Materials Project website.

These technologies were ultimately chosen to achieve the design goal of improving accessibility of the project to newcomers: the Python programming language has seen broad adoption in the scientific community and is now often the first programming language that a new scientist will learn.

Examples on how to get started with Crystal Toolkit to write your own app, including small, self-contained examples and installation instructions, are available in the public documentation\cite{docs}.

\section{Case Study: Powering the Materials Project}

The Materials Project\cite{materialsproject} started as a result of the Materials Genome Initiative\cite{mgi} with a mandate, among other goals, to develop ``open web-based access to computed information on materials and analysis tools to design novel materials.'' To date, it is now in use by over a quarter of a million registered users from academia, industry and government, with a user base that is growing rapidly year-by-year. The Materials Project offers an encyclopedic view of most known, and many hypothetical, inorganic crystalline materials and their predicted properties, with a goal of accelerating the materials discovery process.

\begin{figure*}
    \centering
    \includegraphics[height=0.9\textheight]{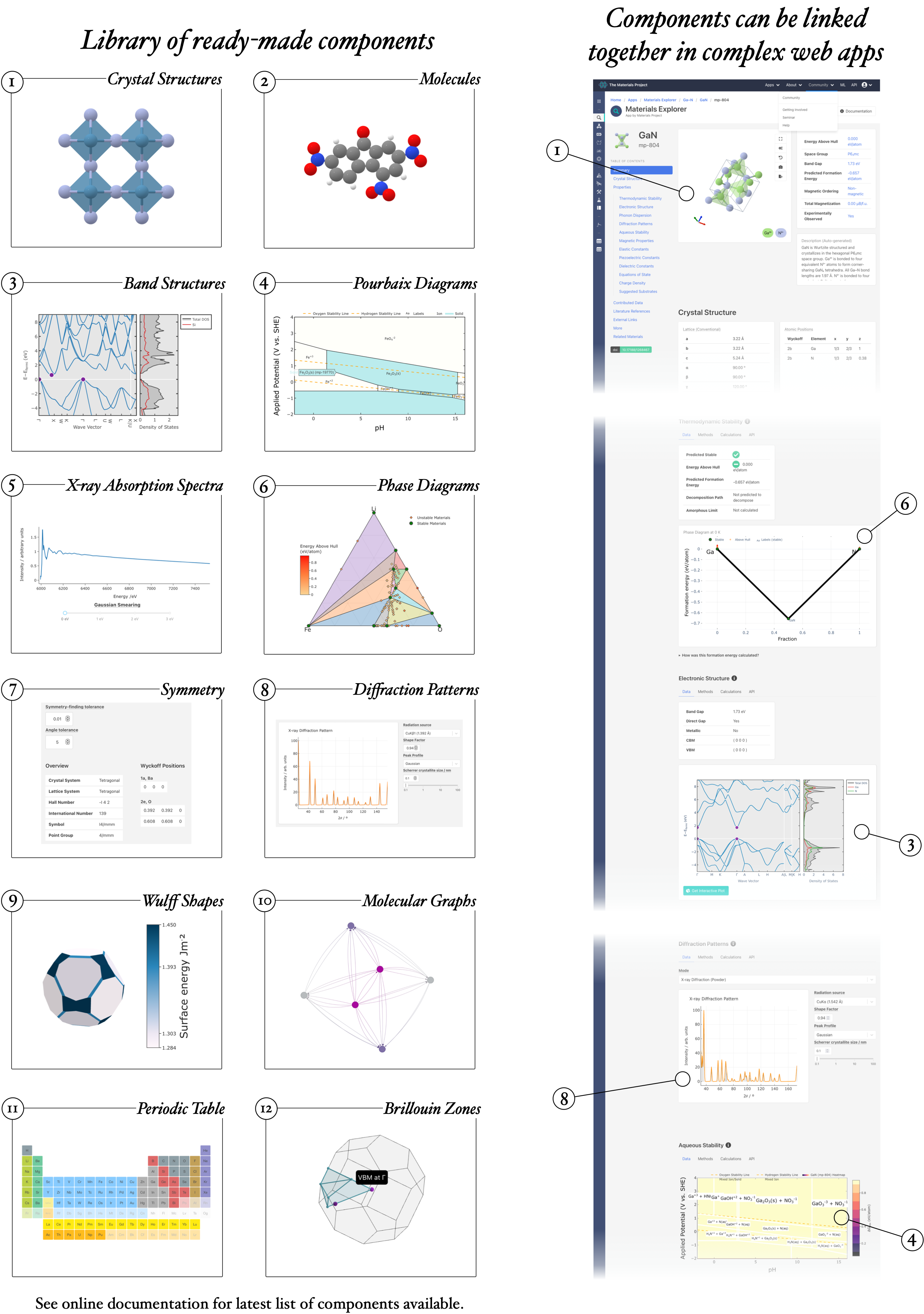}
    \caption{Left, a summary of just some of the components available for use in a Crystal Toolkit web app. Many components rely on underlying functionality available in the \emph{pymatgen} code, and some functionality was specifically added to \emph{pymatgen} to facilitate features in Crystal Toolkit. Right, an example of real-world usage of some of these components in a Materials Project ``materials detail page'', here showing the detail page for wurtzite GaN for this example, located at https://materialsproject.org/materials/mp-804.}
    \label{fig:library}
\end{figure*}

In 2022, Materials Project launched its first major iteration of its web platform (see screenshot in Figure \ref{fig:library}, right panel) based upon the Crystal Toolkit framework. Adopting Crystal Toolkit has allowed more rapid development of new features and apps in Materials Project by opening up development to the broader materials science community including graduate students. For example, a new Catalysis Explorer app, making available information computed information on various adsorbates and crystallographic surfaces, and a new MOF Explorer app\cite{rosen2022high}, making available information on over 20,000 metal--organic frameworks and coordination polymers, have both been made possible by the Crystal Toolkit framework, powered by data supplied by the Materials Project's ``MPContribs'' API.

Crystal Toolkit has also enabled the development of internal tools to examine the Materials Project database as it is being built to include new data, so as to more easily identify issues and perform quality control.

To demo the Crystal Toolkit framework, an app for the Materials Project website has also been developed. The Crystal Toolkit app allows users to upload their own crystal structures and perform transformations using \emph{pymatgen} without having to write their own code. These transformations might include creating a surface slab, assigning oxidation states, or building grain boundaries. The Crystal Toolkit framework began development as an effort to improve this app to include new transformation options, and as the benefit of the approach became evident, the technology grew to encompass the full Materials Project website frontend.

\section{Conclusion}


Crystal Toolkit is a new tool intended for computational materials scientists to help visualize and share their data. It is designed to complement and build upon existing efforts, and offers a modest advance in a few key respects (see ``Design Goals'') by addressing the evolving needs of the materials science community. 

Crystal Toolkit provides ways to visualize many common data types used in materials science, and provides interactive widgets that can be linked together and combined in a web app, as well as providing visualizations in analysis tools such as Jupyter notebooks. Crystal Toolkit is proven at scale, and is now powering Materials Project website, its apps, and visualizations included therein. As a demonstration, images of all 146,000 crystal structures available in Materials Project at time of writing have been generated using Crystal Toolkit, with the intent of releasing images under a Creative Commons license for download on Wikimedia.

In the future, we will look to continue to expand the visualizations available, and the means of interacting with them. Despite advances, in many ways modern computer visualizations are not that dissimilar to Hoffman's original ball-and-stick models. It is important that as new data becomes available that we also reassess the best ways of visualizing this data, and thereby develop new intuitions and find new ways of unlocking the secrets hidden within this data.

\section{Author Contributions}

Authors listed alphabetically with the exception of MKH, JXS, AJ, KAP. MKH initiated and led the project, prepared the manuscript and all figures, developed the code, and is lead maintainer of the code. JXS contributed initial support for Jupyter notebooks, Asymptote rendering, surface plotting, initial support for axes, and various bug fixes, and offered testing and feedback throughout. JB developed the reaction calculator example app. OC developed functionality related to periodic tiling. FC improved the JavaScript code, contributed features to 3D rendering and wrote the Periodic Table React component. AMG contributed a component for Fermi surfaces. RG improved rending of molecules. PH deployed the current Materials Project website based on Crystal Toolkit. HHL contributed functionality for rendering migration graph objects. MM contributed to the phase diagram, X-ray diffraction and X-ray absorption components. J Montoya contributed the Pourbaix component. GM improved functionality related to rendering of crystals with magnetic moments. J Munro contributed band structure component. CTO wrote several custom React components. SZ worked on electron diffraction component, supervised by CO. GP helped with X-ray diffraction component and with general bug fixes and improvements. JR helped with general repository maintenance, continuous integration and bug fixes. SW developed POV-Ray integration. BW helped with developing an example app for display of contributed data via MPContribs and improvements to documentation. DW helped with initial integration within the legacy Materials Project website. RY helped with bug fixes, display of calculation task documents, and absorption components. AJ provided helpful feedback. KAP provided helpful feedback, supervision and funding support.

\section{Acknowledgements}

Shyue Ping Ong is acknowledged for permission to adopt the ``Crystal Toolkit'' name. Shyam Dwaraknath is acknowledged for helpful conversations throughout and contributions to the design. Tyler Huntington is acknowledged for assisting DW in the initial integration of Crystal Toolkit with the Materials Project. Andrew Rosen is acknowledged for feedback and testing of Crystal Toolkit during development of an interactive web app for MOF data and for helpful comments on the manuscript. David Waroquiers is acknowledged for helpful discussions during the development of chemical environment components. Rachel Woods-Robinson is acknowledged for helpful comments on the manuscript. This work was supported by the U.S. Department of Energy, Office of Science, Office of Basic Energy Sciences, Materials Sciences and Engineering Division under Contract No. DE-AC02-05-CH11231 (Materials Project program KC23MP).


\bibliography{iucr}

\end{document}